# Statistical mechanics of base stacking and pairing in DNA melting

Vassili Ivanov, Yan Zeng, and Giovanni Zocchi

*Department of Physics and Astronomy, University of California Los Angeles,*

*Los Angeles, CA 90095-1547*

We propose a statistical mechanics model for DNA melting in which base stacking and pairing are explicitly introduced as distinct degrees of freedom. Unlike previous approaches, this model describes thermal denaturation of DNA secondary structure in the whole experimentally accessible temperature range. Base pairing is described through a zipper model, base stacking through an Ising model. We present experimental data on the unstacking transition, obtained exploiting the observation that at moderately low pH this transition is moved down to experimentally accessible temperatures. These measurements confirm that the Ising model approach is indeed a good description of base stacking. On the other hand, comparison with the experiments points to the limitations of the simple zipper model description of base pairing.

# I. INTRODUCTION

In the DNA double helix, the two strands are held together by the base pairing interaction due to hydrogen bonds between complementary bases on the different strands. Base pairing is stronger for G-C (tree hydrogen bonds; binding free energy $\Delta G_{37}^o \sim 3\,k_B T_{310}$ [1]) than for A-T (two hydrogen bonds; $\Delta G_{37}^o \sim 1.5\,k_B T_{310}$). The other important interaction is between nearest-neighbor bases along the same strand; this is referred to as "base stacking", and induces a partial configurational order even of the single-stranded (ss) molecule. The double helix melts into separate strands upon heating; denaturation of the DNA secondary structure can be detected by the increase in UV absorption. Optical absorption around 260 nm arises from the π-π* electronic transition in both purine and pyrimidine bases. An increase in absorption represents a change in the electronic configuration of the bases; both base unstacking and unpairing contribute to this effect [1]. Indeed, UV absorption of DNA duplex samples continues to rise with temperature after complete separation of the double helix, due to unstacking of the bases in the single strands. Depending on the sequence, the contribution to UV absorption due to unstacking may be comparable to the contribution due to unpairing (see Fig. 1 and 3). For this reason, UV spectroscopy may be advantageous compared to calorimetry for studies of stacking.

Theoretical models are necessary to understand the microscopics of the transition, and also to extract thermodynamic parameters from the experimental measurements [2]. The main approaches are exemplified by the nearest neighbor (NN) thermodynamic model [3], the Poland-Scheraga [4], and the Peyrard-Bishop [5] statistical mechanics models; for short oligomers, the simpler zipper model [6] is sometimes used. Modern developments of Peyrard-Bishop like Hamiltonian models were applied to describe DNA unzipping under an external force [7-10]. Recent theoretical work has concentrated on the nature of the transition in the thermodynamic limit [11-15]; by contrast, here we focus on the question of which degrees of freedom are crucial for a statistical mechanics description. We present experimental data on oligomers which highlight that base pairing and stacking must be considered as independent degrees of freedom; the measurements further show that stacking is well described by an Ising model. Qualitatively, the unpairing transition is relatively narrow in temperature, while stacking interactions display a broader transition (see Figs. 1 and 2). We propose a model where pairing is treated within a zipper model, and stacking within an Ising model approach; the latter was previously suggested for the single stranded (ss) DNA structure in Ref. [2]. We compare this model with experimental measurements for four different oligomer sequences: a sequence of length 60 (L60) designed to form a bubble in the AT rich middle region; a sequence of length 36 designed to unzip from one AT rich end (L36); two short, homogeneous, GC dominated sequence of length 13 (L13 and L13-2). In order to unambiguously pinpoint the critical temperature for strand separation (see arrows in Fig. 1) we measure two different melting curves: the UV absorption curve $f$(T), which monitors a combination of base stacking and pairing, and the dissociation curve $p$(T), which monitors the fraction of completely dissociated molecules, obtained through a method based on quenched states which we recently introduced [16,17]. By design, the model can account for the unstacking contributions to the UV melting curves. To further test the model, we present experimental melting curves of single stranded DNA oligomers, where base stacking are the only contributing degrees of freedom. We find that at moderately low pH (3.6) the unstacking transition is moved below 100 C, so that the whole melting curve is accessible. From these data we show directly that the unstacking transition is well described by the Ising model approach. On the other hand, comparing to the experimental data shows that the zipper model, which was adopted here for simplicity, is deficient in describing the unpairing transition of even short oligomers (see Fig. 3). In summary, this study highlights the role of stacking as a distinct degree of freedom in the statistical mechanics of DNA melting and further demonstrates that an Ising model description is adequate for stacking.

## II. BASE PAIRING

We start from the simple zipper (SZ) model, which describes the UV absorption of double stranded (ds) DNA oligomers up to the temperature of strand dissociation. In this model there is a fixed energy cost and a fixed increase in number of configurations per broken base pair. The latter assumption is unrealistic as it ignores excluded volume effects; the advantage is an elementary analytical solution of the model. With the further assumption that the molecule can only unzip from the ends, the partition function for an oligomer with $N$ base-pairs is given by:

$$Z_{Zipper} = \sum_{p=0}^{N-1}(p+1)\ \exp\left[p(-U/T+\boldsymbol{s})\right] + Z_N,$$

(1)

$$Z_N = \exp\left[N(-U/T+\boldsymbol{s})+\boldsymbol{s}_D\right],$$

where the summation index $p$ is the number of open base-pairs, $Z_N$ describes the state of complete strand separation, $U > 0$ is the pairing enthalpy per base, $\boldsymbol{s} > 0$ is the pairing entropy. The combinatorial factor $(p + 1)$ accounts for unzipping at the *two* ends. Enthalpies and entropies are divided by the Boltzmann constant. The bulk entropy term $\boldsymbol{s}_D$ ("strand dissociation") accounts for the extra entropy gain when the two strands separate ($\boldsymbol{s}_D$ is a function of DNA concentration); it is necessary to correctly describe the dissociation curves $p$ [see Fig. 3 (A2)].

## III. BASE STACKING

The increase in UV absorption after strand separation, visible in Figs. 1 & 3, is due to unstacking in the single strands. The unstacking contribution to the melting curves highlights the necessity of considering base pairing and stacking as separate degrees of freedom. We describe stacking by an Ising model with two parameters: the stacking enthalpy $E$, and the stacking entropy $S$. The probability of a single unstacking at temperature T is then:

$$P(T) = \exp(-E/T+S)/\left[\exp(-E/T+S)+1\right],$$

(2)

and the partition function for ss DNA with at most $s$ unstackings is given by the product of $s$ individual stacking partition functions:

$$Z_{stacking}(s) = \left[1+\exp(-E/T+S)\right]^s.$$

(3)

We consider that a base can be unstacked only if it is unpaired; this is consistent with a zipper model description of pairing. If the N-mer ds DNA has unpaired ends of length $a$ and $b$ ($a+b = p$, $p < N$) we have $2p$ stackings, which might be unstacked. The partition function for the model with pairing and stacking is, from Eqs. (1) and (3):

$$Z_{ZS} = \sum_{p=0}^{N-1} (p+1) \exp[p(-U/T + \boldsymbol{s})] Z_{stacking}(2p) + Z_N$$

$$Z_N = \exp[N(-U/T + \boldsymbol{s}) + \boldsymbol{s}_D] Z_{stacking}(2N-2)$$

(4)

We note that from Eqs. (3) and (4) it is easy to derive the NN model enthalpy and entropy, which are then not constant parameters, but functions of the temperature. Such temperature dependence has been observed in experiments [18,19].

**A. Zipper model for ds DNA CG-rich on one side and AT-rich on the other (ZM2)**

There are two different kinds of bubbles in DNA: bubbles bounded on both sides by ds segments ("bubble-in-the-middle"), and half-bubbles opening from the ends of the molecule ("bubble-at-the-end"). As we have generated experimental data on both kinds [18], here and in the next section we specialize the model to the corresponding sequences.

Let us consider a ds DNA N-mer with a CG - rich region $A$ bases long, and an AT – rich region $B$ bases long, $A + B = N$ (such as the sequence L36). The partition function in our approach is:

$$Z_{ZM2} = Z_N + \sum_{0 \leq a+b < N} \exp[A_1(a) + B_1(b)] Z_{Stacking}[2(a+b)], \quad (5)$$

$$Z_N = \exp[A_1(A) + B_1(B) + \boldsymbol{s}_D] Z_{Stacking}(2N-2). \quad (6)$$

The functions $A_1(x)$, $B_1(x)$ are given below:

$$A_1(x) = \begin{cases} x(-U_{CG}/T + \boldsymbol{s}_{CG}), & x \leq A, \\ A(-U_{CG}/T + \boldsymbol{s}_{CG}) + (x-A)(-U_{AT}/T + \boldsymbol{s}_{AT}), & x > A, \end{cases} \quad (7)$$

$$B_1(x) = \begin{cases} x(-U_{AT}/T + s_{AT}), & x \leq B, \\ B(-U_{AT}/T + s_{AT}) + (x-B)(-U_{CG}/T + s_{CG}), & x > B, \end{cases} \quad (8)$$

where $U_{CG}$, $s_{CG}$, $U_{AT}$, $s_{AT}$ are the pairing enthalpy and entropy (per base) for the CG and AT - rich regions, respectively.

To compare with our experimental data on more complex, non-homogeneous sequences, such as L60, we extended the model in two ways. We introduce different pairing enthalpies and entropies for GC and AT pairing (but maintain, for simplicity, a single entropy and enthalpy of stacking), and we allow in the partition sum states with a single bubble bound by ds tracts. Bubble formation is suppressed by a nucleation cost. For simplicity, we do not explicitly write the partition function here, and instead just show the result in Fig. 1(C).

## RESULTS

For the experiments, synthetic oligomers were annealed as previously described [17]; final concentration for the measurements was 1μM in phosphate buffer saline (PBS) at an ionic strength of 50 mM, pH = 7.4. The sequences used were: **L13** (GCCGCC A GGCGGC), **L13-2** (CGA CGG CGG CGC G), **L36** (CAT AAT ACT TTA TAT T GCC GCG CAC GCG TGC GCG GC) (AT rich at one end), **L60** (CCG CCA GCG GCG TTA TTA CAT TTA ATT CTT AAG TAT TAT AAG TAA TAT GGC CGC TGC GCC) (this has an AT rich tract in the middle), and **PH21** (CGA CGG CGG CGC GCC GTG CGC) (used to study the unstacking transition). The UV absorption $f$ was measured at 260 nm, in a 1 cm optical path cuvette, and normalized so that $f = 1$ corresponds to strand dissociation (see Fig. 1). These and the dissociation measurements have been described before [16, 17, 20]. Briefly, to determine the dissociation curves $p$ (see Fig. 1), samples are heated at temperature $T_i$, then quenched to ~ 0°C. Because the sequences are partially self-complementary, molecules which were dissociated at temperature $T_i$ form hairpins after the quench. The relative number of hairpins (representing the fraction of dissociated molecules $p$ at temperature $T_i$) is determined by gel electrophoresis.

In Fig.1 we display results from the measurements (symbols) and the model (continuous lines). On the left are the UV measurements $f$, on the right the dissociation curves $p$. For the L60 and L36 UV absorption curves, the part of the spectra below 71 and 75 °C, respectively (arrows), corresponds to the melting of the double helix, while the spectra above these temperatures correspond to base unstacking in the ss DNA (compare Figs. 1 and 2). To compare the model with the experimental data, the UV absorption is assumed to be a linear function of pairing and stacking:

$$f = (N\mathbf{a} \times (unpairings) + 2(N-1)\mathbf{d} \times (unstackings) + \mathbf{g})/C \qquad (9)$$

where $C$ is the concentration of the ds oligomer, measured in mmol, $\mathbf{a}$ and $\mathbf{d}$ are the molar extinction coefficients for unpairing and unstacking, measured in mmol$^{-1}$cm$^{-1}$ (the optical path being 1 cm). To fit the model to the data, we used the following procedure. $\mathbf{a}$, $\mathbf{d}$, and $\mathbf{g}$ are found by minimizing the integral of the squared difference between the measured and predicted UV spectra. The unstacking part of the UV spectra was fitted first, using Eq. (3). The stacking enthalpy and entropy were found to be 16000 K and 42 respectively, corresponding to $T_{1/2} = 108^oC$. These high values of the stacking parameters are due to the stronger stacking interactions of the CG tract. AT stackings are weaker [2] and for nonhomogeneous sequences should be considered separately. To obtain initial values for the global fit, the double helix melting part of the UV spectra was fitted first without stacking. Finally a global fit using all thermodynamic parameters was performed in the whole temperature range. The resulting parameters are reported in Table 1. From the table, pairing enthalpies and entropies for A-T are about half the values for G-C. The free energies $U - T\mathbf{s}$ are consistent with literature values (see the Introduction). The strong stacking exhibited by L36 is easily decoupled from pairing using the model. For weak stacking, as in the A-T rich part of L60, it is more difficult to distinguish stacking from pairing. The stacking extinction coefficients $\mathbf{d}$ obtained from the model are in agreement with the extinction coefficients in the literature [1,21]. The increase in pairing extinction coefficient $\mathbf{a}$ between L36 and L60 is due to the increase in the number of weak stackings in the DNA sequence. This deficiency of the model is due to the use of only one set of stacking parameters. The ratios of pairing enthalpies to entropies are for all sequences within 3 % of each other (340 ± 10 K). In the data displayed in Fig.1 we can see only the beginning of the unstacking transition, since the midpoint of unstacking occurs above 100°C. Indeed, according to data in the literature [2, 21] and our own measurements the unstacking transition for many stacking combinations is above 100°C, with the exception of the well studied AA and CC stacking in poly(A) and poly(C) [2]. We found that the unstacking transition temperature can be lowered significantly by lowering the pH. This allows us to investigate directly whether the Ising model gives a good description of the stacking degrees of freedom. In Fig.2 we show *single stranded* unstacking curves obtained at pH = 3.6 for the sequences L13 and PH21. L13 is almost self-complementary, but at this pH the ds structure is not stable at room temperature; PH21 is non-self-complementary, so there is no ds structure at any pH. We see that the unstacking transition is much broader in temperature compared to double helix melting, and is well described by the Ising model with a single parameter set.

The stacking enthalpies obtained from these experiments (Fig.2) were 8770 K and 8000 K, and the stacking entropies 27.3 and 24.6 for the L13 and PH21 oligomers respectively. These are lower than the parameters obtained from the data at pH = 7.4 (Fig.1), showing that the stacking interaction is significantly weakened at low pH.

TABLE 1. Thermodynamic parameters used for the experimental curve fit.

| oligo | $U_{CG}$ | $U_{AT}$ | $s_{CG}$ | $s_{AT}$ | $s_D$ | $a$ | $d$ |
|---|---|---|---|---|---|---|---|
| **L36** | 6640 | 3690 | 19 | 11.05 | 6.4 | 2.28 | 6.98 |
| **L60** | 6690 | 3600 | 19.2 | 10.8 | 12 | 5.30 | 9.30 |
| **L13** | 6950 | N/A | 21 | N/A | 7.2 | 0.73 | 9.98 |
| **L13-2** | 6170 | N/A | 18 | N/A | 0 | 3.77 | 4.49 |

## DISCUSSION

In the NN model, unpairing of the bases and partial unstacking are combined into 10 effective thermodynamic parameters. The advantage is a description in terms of a complete set of parameters for all possible base combinations. However, the NN model considers the influence of stacking on the pairing free energies measured *at the double helix melting temperature*. Thus this model does not describe DNA behavior at temperatures above the double helix dissociation temperature. In contrast in the present model we decouple pairing and stacking into independent degrees of freedom described by different parameters. The need for introducing stacking degrees of freedom is obvious from Fig. 3 (A1), which shows that unstacking contributes a large part to the experimental melting curve. This part is not captured by previous models. The model described here is closest to the Peyrard-Bishop approach [5], which includes a stacking interaction term in the Hamiltonian. It differs in that we describe stacking as an Ising model, which is physically appealing and consistent with the experiments (Fig.2).

The dissociation curves *p* are helpful to unambiguously pinpoint the end point of ds melting; they also reveal the limitations of the zipper model approach, even for short sequences (L13), where bubbles are unimportant. Namely, near the endpoint of ds melting, the experimental dissociation curves are consistently steeper than the model. In fact, the experimental data adumbrate a melting curve without inflection point, characteristic of a discontinuous transition. Other difficulties with the zipper model, apparent in Fig. 3, are discussed later. The physics missing from the zipper-style description is related to the oversimplification of assigning a fixed entropy gain per open base pair.

Hairpin formation, which is not included in the partition function of the model, may contribute to the UV signal at temperatures above the strand dissociation [22]. We examined the stability of the various hairpin structures at different temperatures using the MFOLD server [23]. We conclude that in the case of L60, the contribution of hairpin states above the dissociation temperature is negligible. For L36 there is a small contribution, and for L13 a significant contribution. To further analyze this point, we designed two different 13-mer sequences of similar GC content: L13 and L13-2; the first is self-complementary (forms hairpins), while the second is not. Analyzing secondary structures with the MFOLD server, we expect hairpin formation to give a significant contribution to the UV spectrum beyond strand separation for L13, but no contribution for L13-2. Nevertheless, Fig. 3(B) shows that for L13-2, the increase in UV absorption after strand dissociation (at ~ 73$^o$C) is still significant; this must be attributed to the melting of the remaining stacking structure present in the single strands.

Fig. 3 also highlights the role of the strand dissociation entropy $\boldsymbol{s}_D$ and the more stringent comparison between model and experiment which is possible when the strand dissociation curve *p*(T) is also measured. While it is possible to fit the data in Fig. 1 (A1) without this term, it is not possible to fit the data (A1) *and* the data (A2) without it. The melting transition of L13 is significantly narrower compared to L13-2, due to the competition between the hairpin and duplex formation in L13; for the same reason the dissociation temperature is also lower for L13. The softer melting curve L13-2 can still be fitted by the zipper model, but only with $s_D = 0$ [continuous line in Fig. 3 (B)]. However, introducing an appropriate value for $s_D$ [taken from the fits Fig. 3 (A)] shows that the zipper actually predicts too steep a transition [dashed line Fig. 3 (B)]. In a forthcoming paper we describe an improved model which combines the independent stacking degrees of freedom and Poland's algorithm of bubble counting [24], resolving this difficulty.

In conclusion, we have shown that stacking degrees of freedom must be included in a statistical mechanics description of DNA melting, in order to account for the experimental melting curves in the whole accessible temperature range. Further, we bring direct experimental evidence that these stacking degrees of freedom are well described by an Ising model approach. Thus stacking is uncooperative. In the present work, we used the zipper model to describe pairing, for simplicity. Further work might explore how to incorporate this description of stacking into more realistic Hamiltonians.

## ACKNOWLEDGEMENTS

This research was partly supported by the US-Israel Binational Science Foundation under Grant No. 2000298.

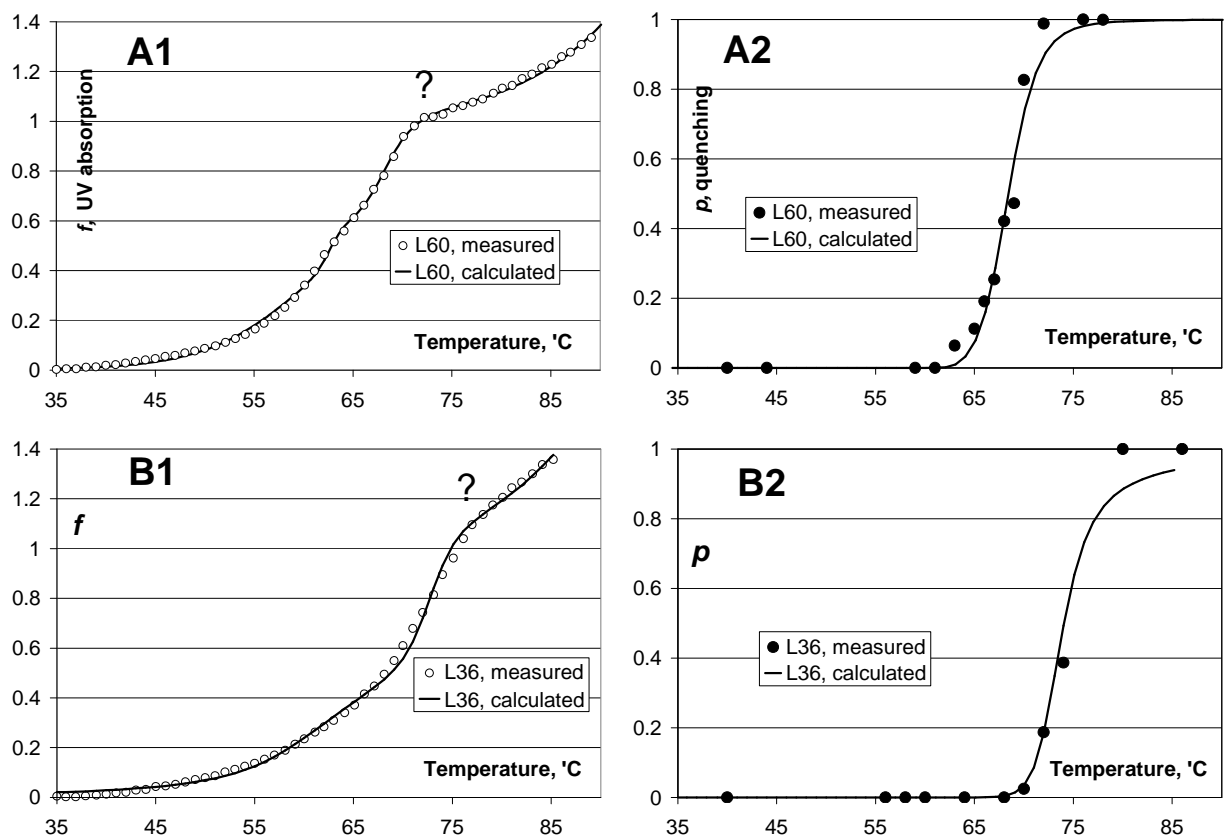

FIG. 1. (A1), (B1). Normalized UV absorption spectra $f$ measured at 260 nm for the L60 and L36 ds DNA oligomers. The experiments are the circles; the model is the solid line. The data for L36 was fitted using eqn. (5-8); for L60 the model was extended to include different pairing interactions for GC and AT, as explained in the text. Arrows show the end point of the ds melting transition, determined from the dissociation curves A2, B2.

(A2), (B2). Measured and predicted dissociation curves $p$, for the same oligomers. The measurements were obtained from the quenching method. The model is plotted using the same parameter values as in (A1), (B1).

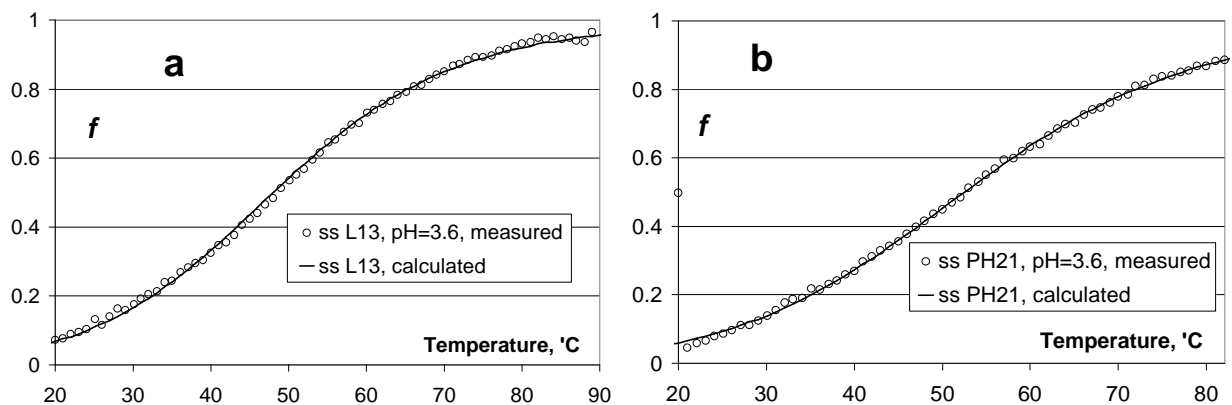

FIG.2. Melting curves obtained from UV absorption for the ss oligomers L13 (a) and PH21 (b). Circles are the experimental data; solid lines are the Ising model eqn. (2). The data were obtained at pH = 3.6 , in order to lower the midpoint of the unstacking transition.

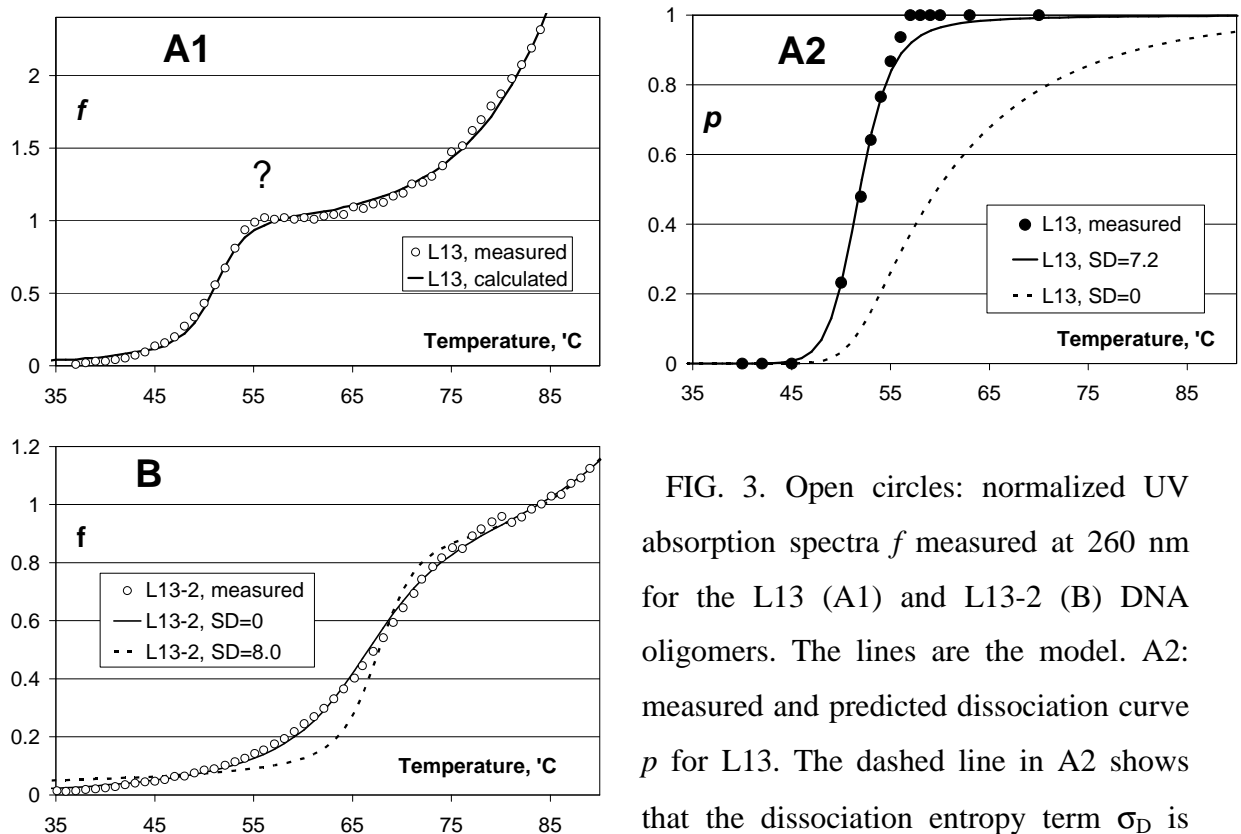

FIG. 3. Open circles: normalized UV absorption spectra $f$ measured at 260 nm for the L13 (A1) and L13-2 (B) DNA oligomers. The lines are the model. A2: measured and predicted dissociation curve $p$ for L13. The dashed line in A2 shows that the dissociation entropy term $\sigma_D$ is needed to account for the $p$ curves. The continuous line in B is a fit using the model with $\sigma_D = 0$ ($s_D$ is called SD in the figures). The dashed line in B shows that the UV spectrum of L13-2 cannot be fitted with the appropriate value $s_D = 8.0$. This deficiency is due to the oversimplified zipper model description.

[1] V.A. Bloomfield, D.M. Crothers, and I. Tinoco, Jr., *Nucleic Acids, Structures, Properties, and Functions*, (University Science Books, Sausalito, CA, 2000).

[2] S.M. Freier, *et al*., Biochemistry **20**, 1419 (1981), and references there in.

[3] J. SantaLucia, Jr., Proc. Natl. Acad. Sci. USA **95**, 1460 (1998).

[4] D. Poland and H.A. Scheraga, J. Chem. Phys. **45**, 1456 (1966); **45**, 1464 (1966).

[5] M. Peyrard and A.R. Bishop, Phys. Rev. Lett. **62**, 2755 (1989); T. Dauxois, M. Peyrard, and A.R. Bishop, Phys. Rev. E **47**, 684 (1993).

[6] C. Kittel, Am. J. of Phys. **37**, 917 (1969).

[7] D.K. Lubensky and D.R. Nelson, PRL, **85**, 1572 (2000).

[8] D. Marenduzzo, S. M. Bhattacharjee, A. Maritan, E. Orlandini, and F. Seno, Phys. Rev. Lett. 88, 028102 (2002).

[9] D. Marenduzzo, A. Trovato, and A. Maritan, Phys. Rev. E 64, 031901 (2001).

[10] C. Danilowicz, V.W. Coljee, C. Bouzigues, D.K. Lubensky, D.R. Nelson, and M. Prentiss, PNAS **100**, 1694 (2003).

[11] N. Theodorakopoulos, T. Dauxois, and M. Peyrard, Phys. Rev. Lett. **85**, 6 (2000).

[12] Y. Kafri, D. Mukamel, and L. Peliti, Phys. Rev. Lett. **85**, 4988 (2000).

[13] M.S. Causo, B. Coluzzi, and P. Grassberger, Phys. Rev. E **62**, 3958 (2000).

[14] D. Cule and T. Hwa, Phys. Rev. Lett. **79**, 2375 (2000).

[15] E. Carlon, E. Orlandini, and A.L. Stella, Phys. Rev. Lett. **88**, 198101 (2002).

[16] A. Montrichok, G. Gruner, and G. Zocchi, Europhys. Lett. **85,** 452 (2003).

[17] Y. Zeng, A. Montrichok, and G. Zocchi, Phys. Rev. Lett. **91**, 148101 (2003).

[18] P. Wu, S. Nakano and N. Sugimoto, Eur. J. Biochem. **269,** 2821 (2002).

[19] I. Rouzina and V.A. Bloomfield, Biophys. J. **77**, 3242 (1999); J. **77**, 3252 (1999); Biophys. J. **80**, 882, 894 (2001); **80**, 894 (2001).

[20] Y. Zeng, A. Montrichok, and G. Zocchi, J. Mol. Biol. **339**, 67 (2004).

[21] C.R. Cantor and P.R. Schimmel, *Biophysical Chemistry, Part I: The conformation of biological macromolecules*. (W.H Freeman and Company, New York, 1999), p.328

[22] Yu.L. Lyubchenko (private communication_, see also D.B. Naritsin and Yu.L. Lyubchenko, J. of Biomol. Struct. Dyn., **4**, 813 (1991).

[23] M. Zuker, Nucl. Acids Res. **31**, 3406, (2003); D.H. Mathews, J. Sabina, M. Zuker and D.H. Turner J. Mol. Biol. **288**, 911 (1999)

[24] D. Poland, Biopolymers **13**, 1859 (1974).